\documentclass{moriond}
\usepackage{latexsym}
\usepackage{graphicx}
\usepackage{color}
\usepackage{amsmath}
\usepackage{amssymb}

\def\be{\begin{equation}}
\def\ee{\end{equation}}
\def\bea{\begin{eqnarray}}
\def\eea{\end{eqnarray}}

\def\good{\raisebox{0.35mm}{{\color{green}$\bigstar$}}}
\def\soso{\hspace{0.25mm}\raisebox{-0.2mm}{{\color{green}\Large$\circ$}}}
\def\bad{\raisebox{0.35mm}{\hspace{0.65mm}{\color{red}\tiny$\blacksquare$}}}

\newcommand{\Photo}{\includegraphics[height=35mm]{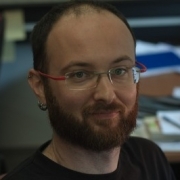}}

\begin{document}
\vspace*{4cm}
\title{LATTICE INPUTS TO FLAVOR PHYSICS}

\author{ M. DELLA MORTE }

\address{CP$^3$-Origins and Danish IAS, Campusvej 55, 5230 Odense M, Denmark 
and \\ IFIC (CSIC), Calle Catedr\'atico Jos\'e Beltran, 2, 46980 Paterna, Valencia, Spain
}

\maketitle\abstracts{
We review recent lattice results for quark masses and low-energy hadronic parameters
relevant for flavor physics. We do that by describing the FLAG initiative, with emphasis
on its scope and rating criteria. The emerging picture is that while for {\it light} quantities a large number
of computations using different approaches exist, and this increases the overall confidence on the final
averages/estimates, in the {\it heavy-light} case the field is less advanced and, with the exception of
decay constants, only a few computations are available.\\
The precision reached for the {\it light} quantities is such that electromagnetic (EM) corrections, beyond the point-like
approximation, are becoming relevant. We discuss recent computations of the spectrum based on direct simulations of QED+QCD. We also
present theoretical developments for including EM effects in leptonic decays.\\
We conclude describing recent results for the $K \to \pi \pi$ transition amplitudes and prospects for tackling hadronic decays on the lattice.
}

\section{Introduction and FLAG}
After its discovery in 2012, the Higgs boson was believed to provide a portal to New Physics. This is even somehow assumed when 
formulating the hierarchy problem of the Standard Model (SM).
However, this far, all measurements of the Higgs boson properties lie within 20\% of the SM expectations, as reported 
by ATLAS~\cite{ATLAS-h} and CMS~\cite{CMS-h}.
Instead, there is a number of 2-3 sigmas tensions in rare processes (see for example~\cite{AltStau}), 
with the most prominent examples being in the angular analysis of the $B^0 \to K^{*0} \mu^+ \mu^-$ decay~\cite{LHCb-BK}
and in the enhancement of the $B \to D^{(*)} \tau \bar{\nu}_\tau$ decays~\cite{Lees:2012xj}
.
Significances depend on treatment of several non-perturbative effects.
Extrapolating to the future, (some of) these rare decays won't be so rare anymore.
Belle~2 will report results from about 2018 and coexist with
the LHC and High Luminosity (HL)-LHC, after Long Shutdown 3 in 2023-2025. Progress on the theoretical side is needed in many instances, to
match the expected experimental accuracy.

There are many different groups, all over the world, using different lattice 
methods, that calculate hadronic matrix elements relevant for a number of weak decay processes of 
$K$, $D_{(s)}$, and $B_{(s)}$ mesons. 
With so many groups calculating similar matrix elements using different methods, and
all providing phenomenologically relevant results with complete error budgets,
 it is useful to try to produce global averages/estimates and to review virtues and shortcomings
of the different computations in a transparent way, which should be accessible also to the non-experts.
This is the goal of the FLAG initiative.

\subsection{The FLAG review}
The Flavor Lattice Averaging Group started its activity in 2010 focusing on light-quark quantities and providing 
averages from lattice results with comprehensive error budgets~\cite{Colangelo:2010et}. A second similar initiative was started at around the same 
time~\cite{Laiho:2009eu}, focusing on both heavy- and light-quark quantities. The two groups joined for the second edition of the 
FLAG-review~\cite{Aoki:2013ldr} (FLAG-2). One of the main goals of FLAG is to assess the reliability of systematic error estimates, in particular
concerning continuum extrapolations, chiral extrapolations, finite volume effects and renormalization. This is done through quality
criteria by assigning to each computation a symbol for each one of the systematics above. For example, the symbols and the criteria 
adopted for the light-quark quantities in FLAG-2 are:
\begin{itemize}
\item Chiral extrapolation:\\
\good \hspace{0.14cm}  $M_{\pi,\mathrm{min}}< 200$ MeV  \\
\rule{0.05em}{0em}\soso \hspace{0.2cm}  200 MeV $\le M_{\pi,{\mathrm{min}}} \le$ 400 MeV \\
\rule{0.05em}{0em}\bad \hspace{0.2cm}  400 MeV $ < M_{\pi,\mathrm{min}}$ \\
in addition it is assumed that the chiral extrapolation is done using at least three points.
\item Continuum extrapolation:\\
\good \hspace{0.14cm}  3 or more lattice spacings, at least 2 points below 0.1 fm\\ 
\rule{0.05em}{0em}\soso \hspace{0.2cm}  2 or more lattice spacings, at least 1 point below 0.1 fm \\ 
\rule{0.05em}{0em}\bad \hspace{0.2cm}  otherwise\\
in addition it is assumed that the action is $O(a)$-improved (i.e. the
discretization errors vanish quadratically with the lattice spacing).
\item Finite-volume effects:\\
\good \hspace{0.14cm}  $M_{\pi,\mathrm{min}} L > 4$ or at least 3 volumes \\
\rule{0.05em}{0em}\soso \hspace{0.2cm}  $M_{\pi,\mathrm{min}} L > 3$ and at least 2 volumes \\
\rule{0.05em}{0em}\bad \hspace{0.2cm}  otherwise.
\item Renormalization (where applicable):\\
\good \hspace{0.14cm}  non-perturbative\\
\rule{0.05em}{0em}\soso \hspace{0.2cm}  1-loop perturbation theory or higher  with a reasonable estimate of truncation errors\\
\rule{0.05em}{0em}\bad \hspace{0.2cm}  otherwise. 
\end{itemize}
For heavy-light quantities the criteria are similar, with some additional ones concerning discretization effects and treatment of heavy quarks.
In general criteria are expected to change in time and possibly become stricter as lattice computations reach new levels of accuracy.
In the end, all the {\it published} (in journals) results with no red symbols enter the final estimates/averages.
In some cases, the averaging procedure leads to results which in the opinion of the authors
do not cover all uncertainties. In these cases, in order to stay on the conservative
side, averages are replaced by estimates (or ranges), which are considered  fair
assessments of the current knowledge acquired on the lattice. 
These estimates are based on a critical (and to some extent subjective) analysis of the available
information.

In detail, the FLAG-2 collaboration counted 28 members representing the major lattice groups in the world.
Different Working Groups were in charge of reviewing different sets of quantities:
Quark masses (WG1), $V_{us},V_{ud}$ (WG2), $\chi$PT Low Energy Constants (WG3), $B_K$ (WG4), 
$f_{B_{(s)}}$, $f_{D_{(s)}}$, $B_B$ (WG5, our group), $B_{(s)}$, $D$ semileptonic and radiative decays (WG6),
and finally $\alpha_s$ (WG6).
In the following we will focus on the subset of quantities presented during the talk.
A more recent update of lattice results concerning heavy-light quantities can be found in~\cite{Bouchard:2015pda}.
In Fig.~1
we show the results for the strange quark mass and the average up and down quark mass.
\begin{figure}[t]
\hspace{-1.2cm}
\includegraphics[width=9.cm]{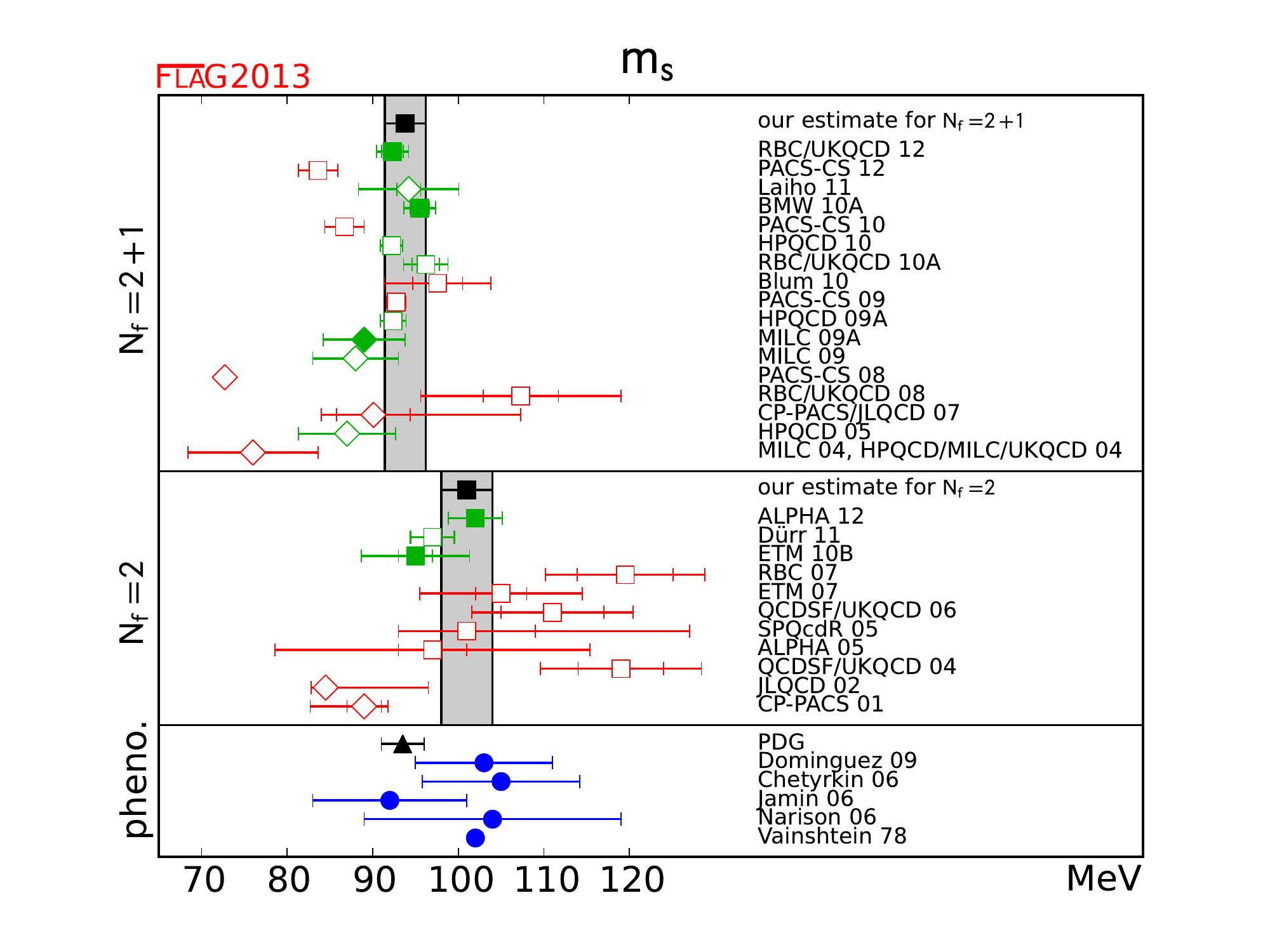}
\includegraphics[width=9.cm]{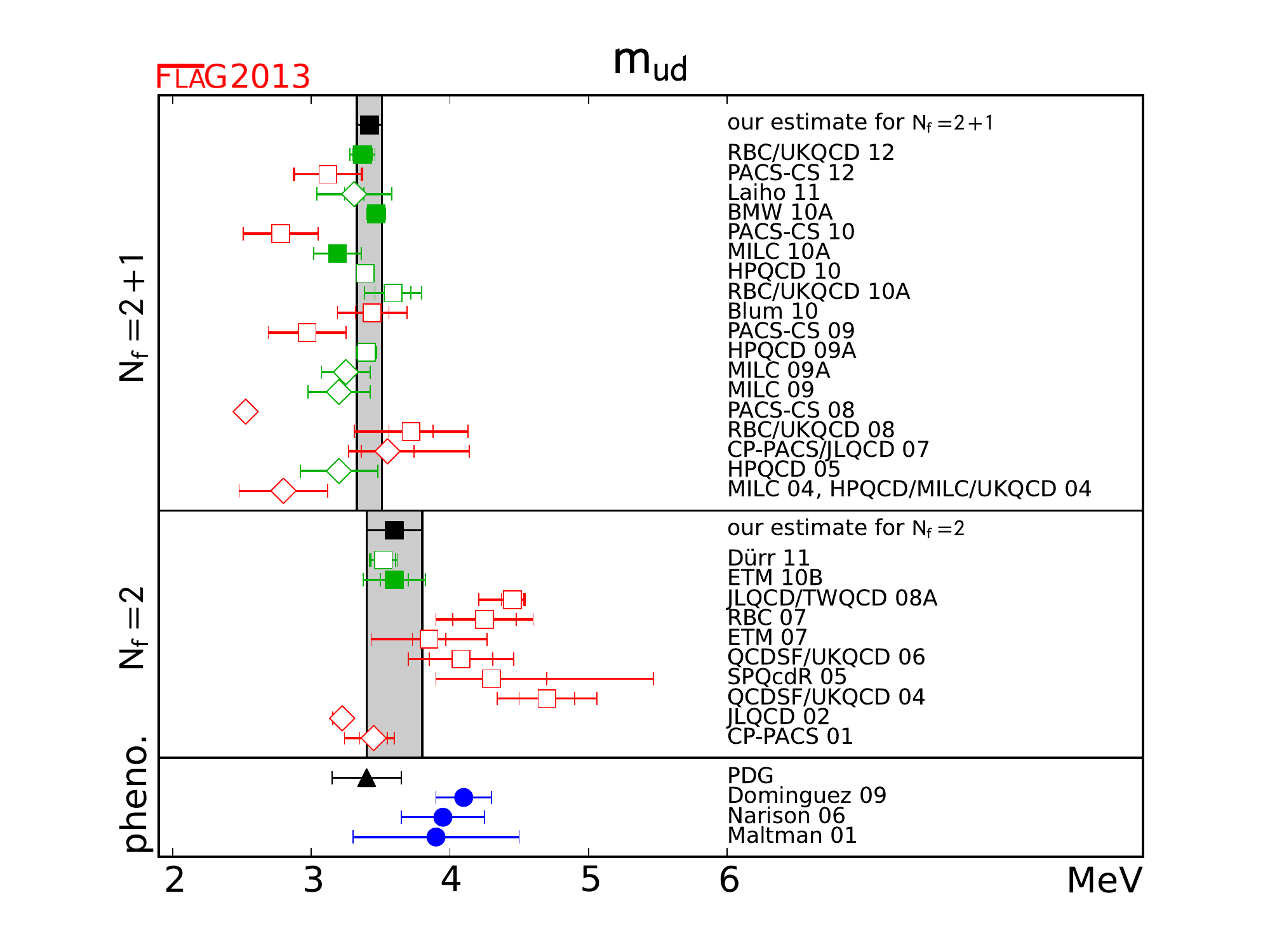}
\label{fig:quarkmasses}
\vspace{-0.35cm}
\caption{Lattice results for the strange quark mass $m_s$ and the average up and down quark mass in the $\overline{\rm MS}$ scheme at the 2 GeV scale.
The bottom panels represent non-lattice (and PDG) results. The FLAG-2 final estimates, including filled green points only, are given by the grey bands.
Figure from$\,^8$.}
\end{figure}
That also serves the purpose of clarifying the difference between averages and estimates discussed above. Indeed, in the $N_f=2+1$ case an error
has been included in the final estimates accounting for the quenching of the charm quark (see~\cite{Aoki:2013ldr} for details).

A second instructive example is taken from the $V_{us},V_{ud}$ working group. These CKM matrix element can be extracted from 
leptonic as well as semileptonic decays and therefore the WG2 focuses on kaon and pion decay constants as well as on form-factors relevant
for the $K \to \pi \ell \nu$ transition. In particular the form-factor $f_+(0)$ at zero momentum transfer is relevant for phenomenology 
and for comparisons to $\chi$PT.  
In Fig.~2 
we show the summary plots for these quantities from~\cite{Aoki:2013ldr}.
\begin{figure}[hb]
\hspace{-1.2cm}
\includegraphics[width=9.cm]{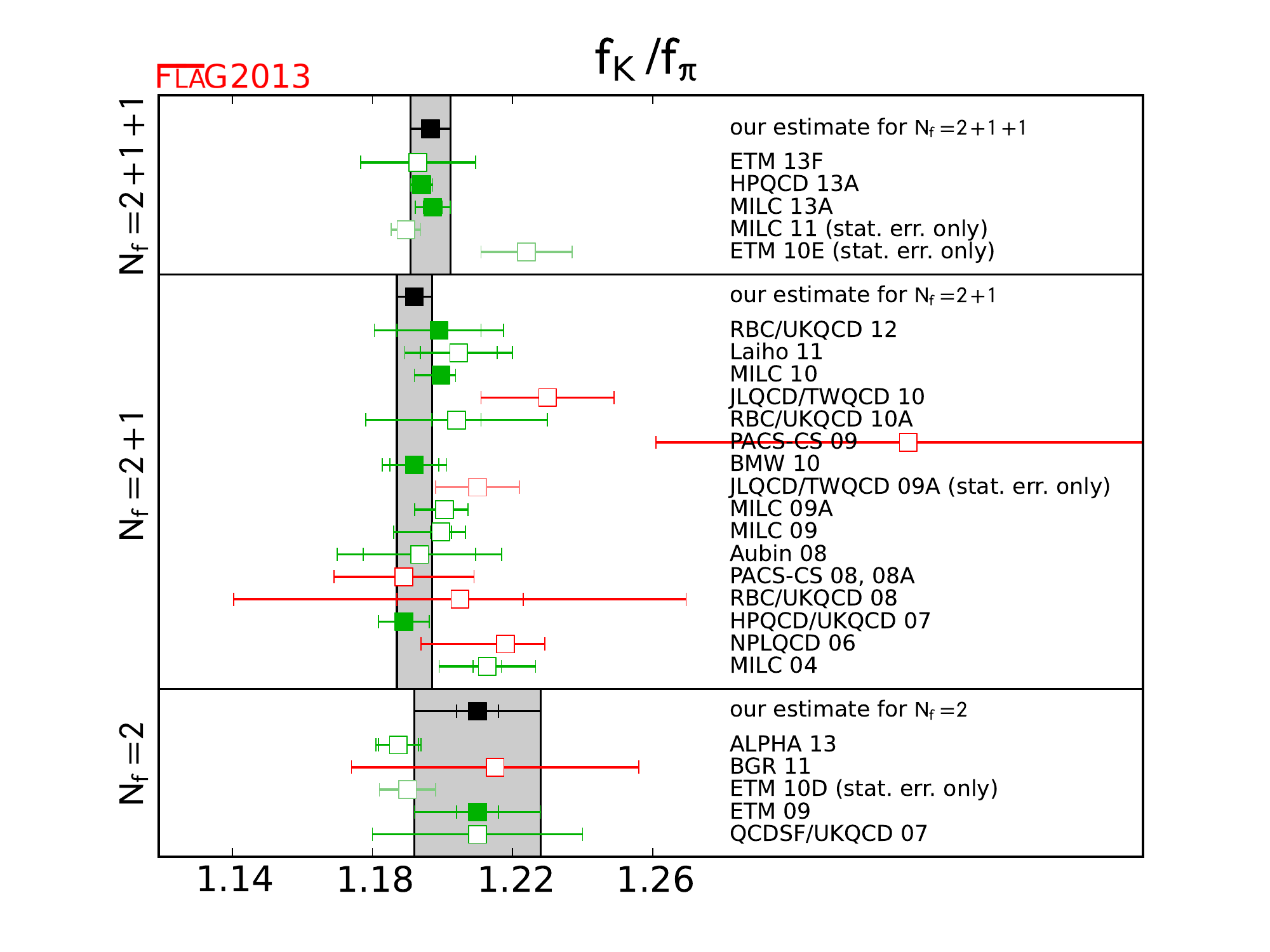}
\includegraphics[width=9.cm]{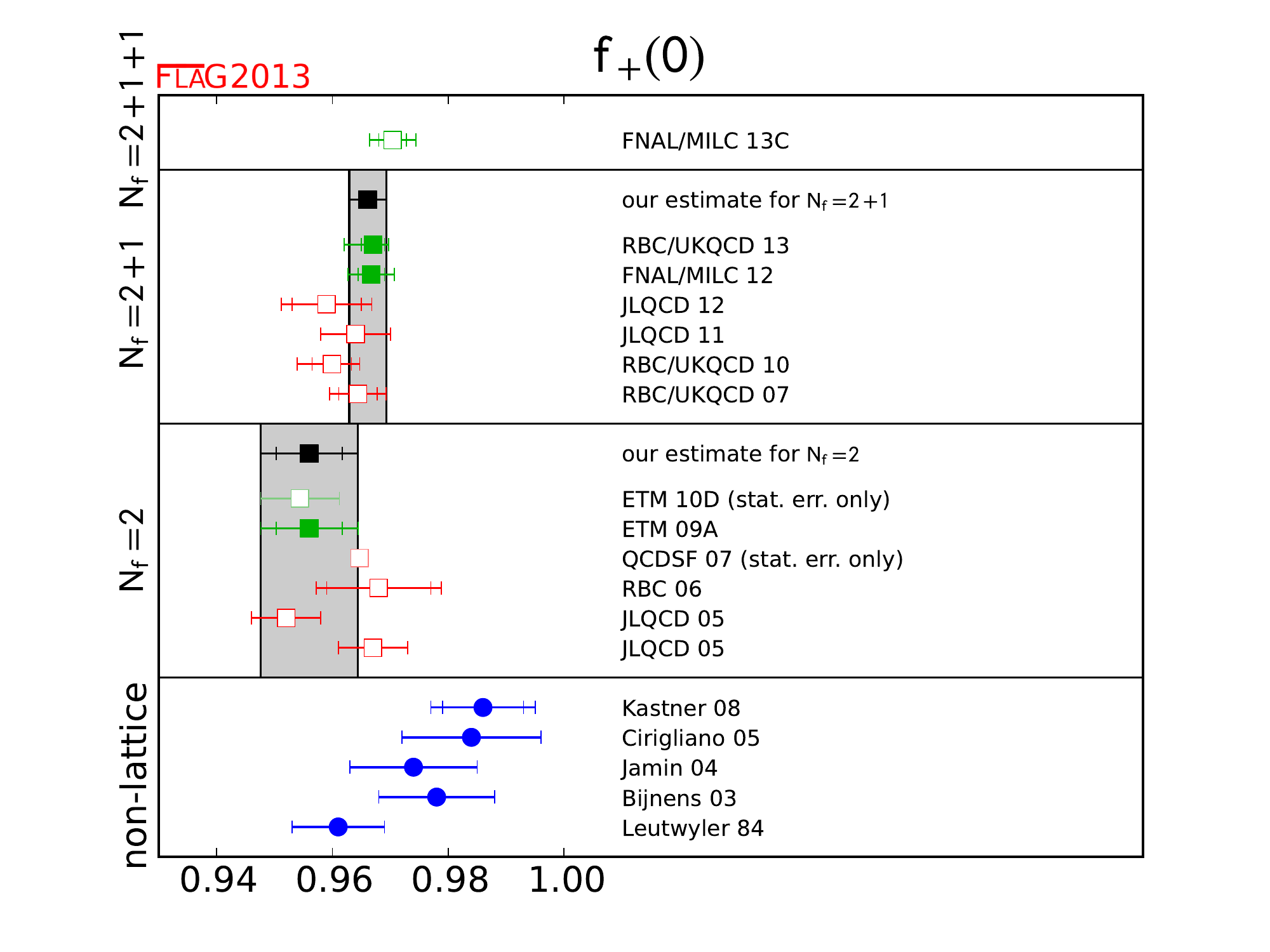}
\label{fig:kl2kl3}
\vspace{-0.35cm}
\caption{Lattice results for $f_+(0)$ and the ratio of decay constant $f_\pi/f_k$. 
The FLAG-2 final estimates, including filled green points only, are given by the grey bands. The blue points are non-lattice estimates.
Figure from$\,^8$.}
\end{figure}
The results can be used to check the first row unitarity of the CKM matrix in the SM. Neglecting $V_{ub}$, the $N_f=2+1$ estimates give
$|V_{ud}|^2 + |V_{us}|^2=0.987(10)$. In addition, as discussed in the review, the consistency of leptonic and semi-leptonic
determinations of $|V_{us}|$ is a check of the equality of the Fermi constant
describing interactions among leptons, and the one describing
interactions among leptons and quarks. This gives an important constraint on possible
modifications and extensions of the SM.

As mentioned, the current situation in the heavy-light sector is much less satisfactory. While some  quantities like 
decay constants have been computed by a number of collaborations using a large variety of methods, for more complicated
ones like form-factors, even for the ``simplest'' pseudoscalar to pseudoscalar, tree-level induced, semileptonic transitions, only
a few determinations exist. In Fig.~3
we show the FLAG-2 summaries for $f_{B_{(s)}}$ and for the
$B \to \pi \ell \nu$ form factor $f_+(q^2)$. For the latter only two computations, based on the same ensembles of configurations
(but using different treatments of heavy quarks) exist. The situation 
is similar for other quantities like the $B_{B_{(s)}}$ mixing parameters (see~\cite{Aoki:2013ldr}).
Very much like experimental results, the confidence increases when 
several results from different collaborations/experiments become available. 
Significant progress in this directions is indeed expected from the lattice community in the next few years and will be visible already in the next FLAG review (expected for 2016).
\begin{figure}[h!]
\vspace{-0.35cm}
\hspace{-1.2cm}
\includegraphics[width=9.cm]{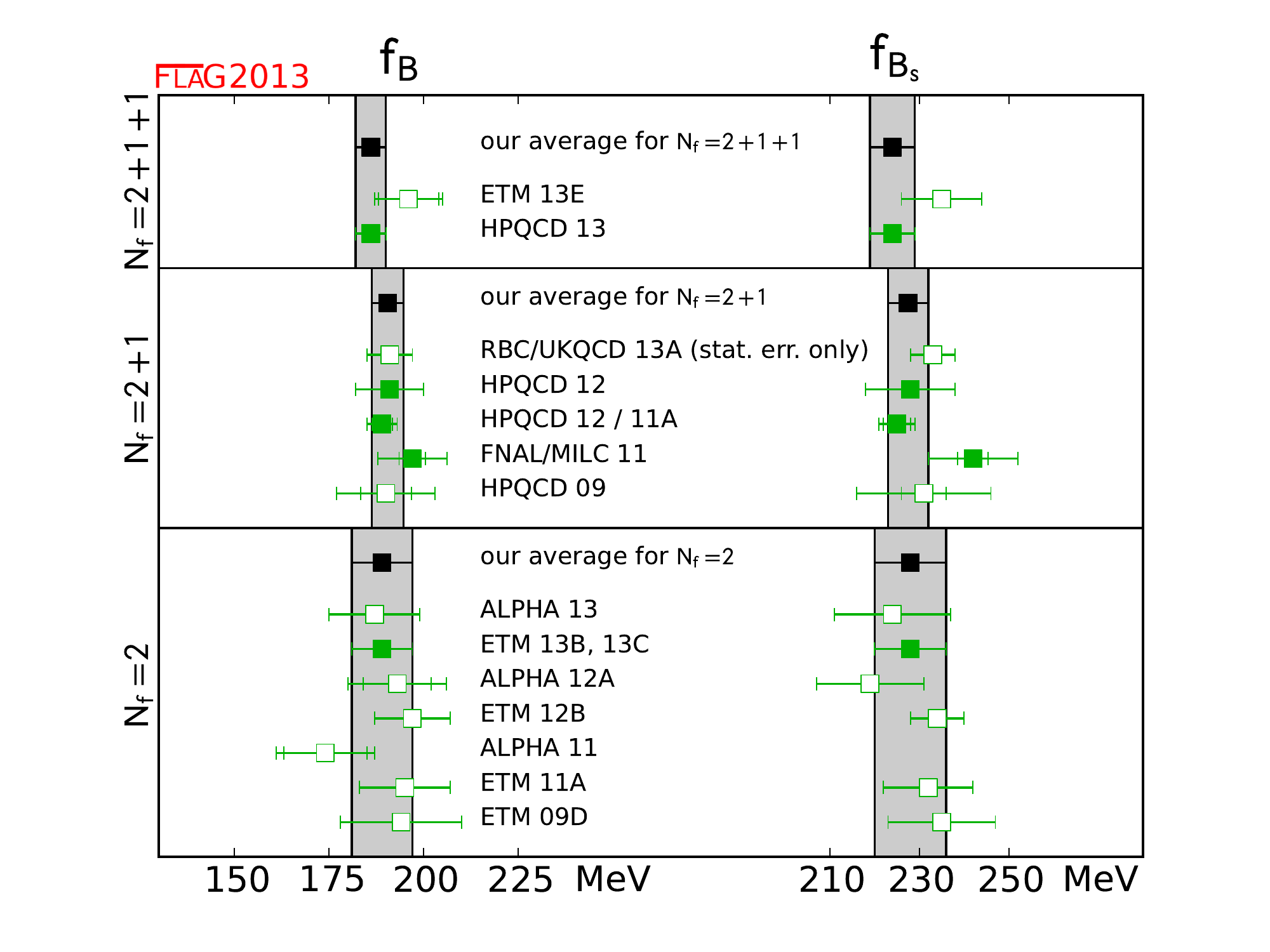}
\hspace{-0.7cm}
\includegraphics[width=8.6cm]{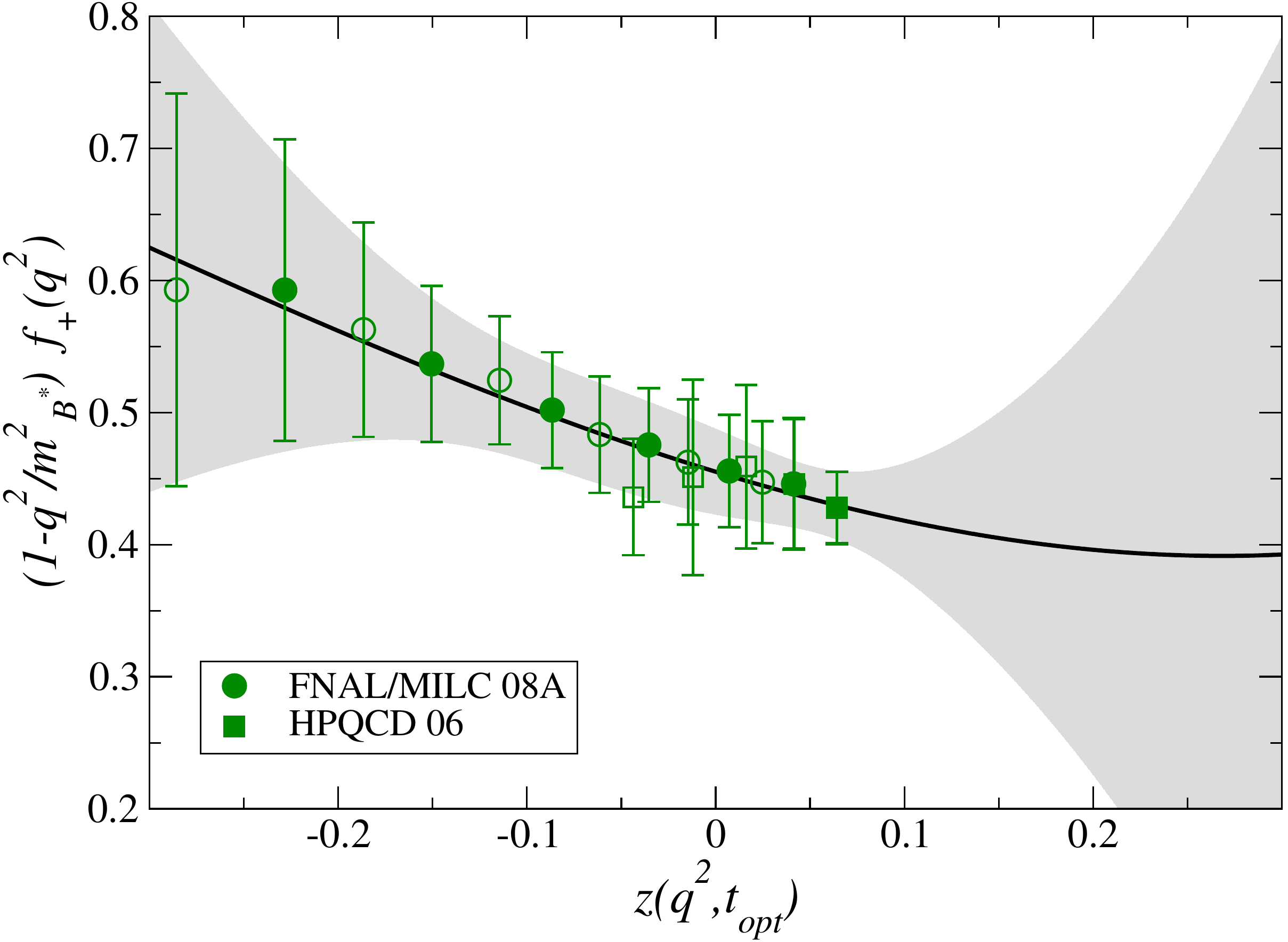}
\label{fig:heavyFLAG}
\vspace{-0.35cm}
\caption{Lattice results for $f_{B_{(s)}}$ (left) and for the
$B \to \pi \ell \nu$ form factor $f_+(q^2)$ (right). 
The BCL parameterization$\,^9$ is used for the latter. The form-factor is expressed as a function of the
$z$ variable, obtained from $q^2$ through a conformal transformation depending on a real parameter $t_{opt}$.
Figure from$\,^8$.}
\end{figure}
\section{Inclusion of EM interactions in lattice QCD computations}
Most of the lattice calculations concerning the
properties of the light mesons are performed in the isospin limit of QCD and neglecting EM interactions.
However, at the precision reached (e.g., the FLAG-2 estimates for the pion and kaon decay constants have an error $ \leq 1\%$), 
they cannot be ignored anymore. For example, the EM corrections to the mass of the charged pions are estimated to be 4~-~5 MeV.
The current approach mostly relies on
$\chi$PT for correcting lattice data in order to include both EM and strong isospin breaking effects. Obviously
it would be desirable to deal with the corresponding terms directly at the level of the simulations.

The BMW Collaboration reported in~\cite{Borsanyi:2014jba} about  
the first direct simulations of QED and QCD with
four non-degenerate flavors, in a fully dynamical formulation. The goal
is the first-principle computation of the neutron-proton mass difference,
a tiny (0.14\%) effect, which is crucial in explaining
the Universe as we know it. This impressive computation is summarized in Fig.~4,
where the results for
the contour lines of the neutron-proton mass difference are given in terms of the $m_u-m_d= \delta m$ splitting
and the EM coupling $\alpha$ (both normalized to their physical value)~\footnote{The separation among EM and strong (QCD) isospin breaking
effects is ambiguous by O($\alpha \delta m$). The prescription adopted in the LO (in isospin breaking) calculation in$\,^{11}$ fixes the EM correction to 
the mass difference $m_{\Sigma^-} - m_{\Sigma^+}$ to vanish.}.
Within the same approach the authors of~\cite{Borsanyi:2014jba} also compute the mass splittings in the $\Sigma$, $\Xi$, $\Xi_{cc}$ and $D$ channels.
They also provide an estimate of the numerical cost for such a computation. Considering the various extrapolations/interpolations in masses and couplings,
the poor statistical signal for small values of the electromagnetic constant and the need for very large volumes, such a calculation turned
out to be 300 times more expensive then their pure QCD computation of the spectrum of stable hadrons in the theory with two dynamical flavors.
\begin{figure}[htb]
\begin{center}
\vspace{-0.1cm}
\includegraphics[width=9.cm]{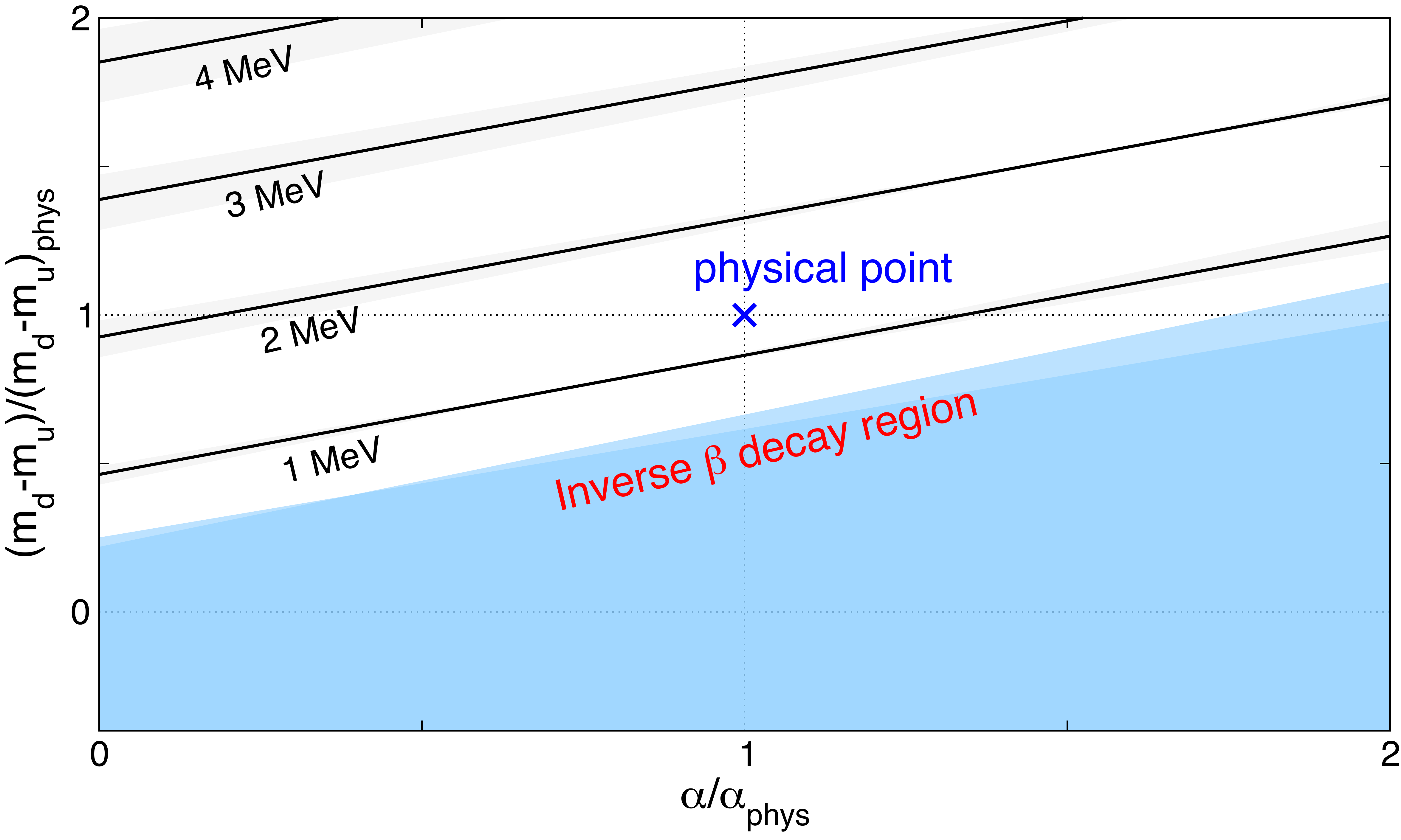}
\end{center}
\label{fig:BMWQED}
\vspace{-0.35cm}
\caption{Contour lines for the neutron-proton mass difference resulting
from a direct QED+QCD computation with 1+1+1+1 (i.e., non-degenerate) dynamical flavors.
Figure from$\,^{11}$.}
\end{figure}

Indeed finite volume effects are one of the main issue in simulating QED on a lattice because of the long-range nature of the EM interactions.
In particular in a finite volume with periodic boundary conditions (in space) zero modes of the gauge field exist, which can not be eliminated through
(standard) gauge-fixing conditions.
In~\cite{Borsanyi:2014jba} the finite-volume zero mode is removed through a non-local constraint. In fact, a rigorous, all-order, proof of the renormalizability 
of the theory in this setup is still missing. An alternative could be to give a mass to the photon perhaps \`a la Stueckelberg~\cite{Ruegg:2003ps}~(and references therein). 
The massless limit (which would have to be taken numerically) is smooth in this case, at least in the continuum.
On a similar line, it may be interesting to reconsider soft covariant gauges, as proposed and studied in~\cite{Henty:1996kv} for non-Abelian gauge theories. 

Additional issues, due to infrared divergences, must be taken into account when trying to include QED corrections into the computation of transition amplitudes. 
These are already present in the case
of the decay constants (or better, the case of leptonic decays), as discussed in~\cite{Carrasco:2015xwa}. 
Let us consider the widths describing the  $\pi^+ \to \ell^+ \nu$ decay
at O($\alpha$) and label them as $\Gamma_i$,
with $i$ the number of photons in the final state. 
It is well known that to obtain physical quantities radiative corrections from virtual and real photons must be combined.
Therefore, at this order we are interested in $\Gamma_0$ and $\Gamma_1(\Delta E)$, where the energy of the photon in the final
state, and in the rest frame of the $\pi^+$, is integrated from $0$ to $\Delta E$.
For the sake of the argument on which the approach in~\cite{Carrasco:2015xwa} is based, it is sufficient to look at the subset of diagrams shown in Fig.~5,
all contributing to $\Gamma_0$. The first (from the left) gives the pure QCD contribution and it is factorizable into an hadronic part (encoded in the matrix element
of the axial current between the vacuum and a $\pi^+$, i.e., the decay constant) and a leptonic one, because a $W$ boson is exchanged in between the two vertices.
The second one is again factorizable and could be viewed as an O($\alpha$) correction to the decay constant, however it is infrared divergent. These divergences
are removed by considering diagrams as the rightmost one and diagrams where a photon is emitted either from a quark line or the lepton line (diagrams contributing to
$\Gamma_1$). But the third diagram, where a photon is exchanged between a quark and the lepton, is not factorizable, so there is really not much physical sense 
in ``QED corrections to decay constants'', rather one should consider corrections to the whole transition process.\\
%
\begin{figure}[htb]
\hspace{-.1cm}
\includegraphics[width=5.1cm]{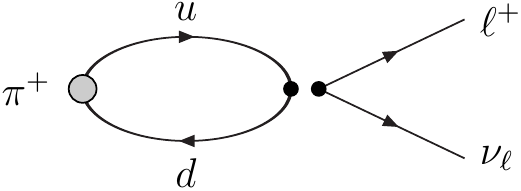}
\includegraphics[width=5.35cm]{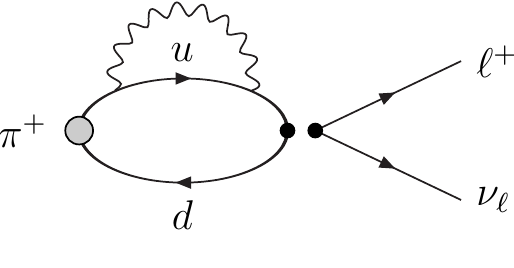}
\hspace{-0.1cm}
\includegraphics[width=5.35cm]{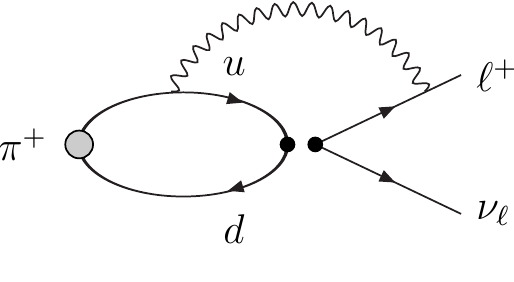}
\label{fig:pilnu}
\vspace{-0.35cm}
\caption{Pure QCD (left) and examples of factorizable (middle) and non-factorizable (right) contributions to $\Gamma_0$ at O($\alpha$), see text.
Figure from$\,^{14}$.}
\end{figure}
In principle, both $\Gamma_0$ and $\Gamma_1(\Delta E)$ could be computed on the lattice, however the latter would be computationally very expensive. Instead, the 
authors of~\cite{Carrasco:2015xwa} propose to use the pointlike ($pt$ in the following formulae) approximation to calculate  $\Gamma_1(\Delta E)$. Values of 
$\Delta E$ around 10~-~20 MeV are experimentally accessible and for such soft photons the coupling to hadrons is conceivably well described by the pointlike
approximation\footnote{The Lagrangian describing the interaction of a pointlike meson with the leptons is non-renormalizable by power counting, very much
like the chiral Lagrangian, which could indeed have also been used here (although more complicated from the analytical point of view when 
considered in a finite volume). For the process described, however, these interactions are inserted at tree-level only and therefore there is no need
for additional counterterms. The only requirement for the method to work is that the contributions from small momenta are the same in the full theory and in its
 approximation. Still, for larger values of $\Delta E$ some approximations may be more accurate than others.}.
In order to ensure an accurate cancellation of the infrared divergences, and since $\Gamma_0$ is computed on the lattice
through numerical simulations,  whereas $\Gamma_1(\Delta E)$ is 
computed in perturbation theory (and in the $pt$-approximation), it is convenient to introduce an intermediate step and re-write
\begin{eqnarray}
\Gamma(\Delta E) &\equiv&  \Gamma_0+ \Gamma_1(\Delta E) = \left\{\Gamma_0 - \Gamma_0^{pt}   \right\} + \left\{\Gamma_0^{pt} + \Gamma_1^{pt}(\Delta E)   \right\}  \nonumber \\
&=& \lim_{L \to \infty} \left\{\Gamma_0(L) - \Gamma_0^{pt}(L)   \right\} \;+\; \left\{  \Gamma_0^{pt}+ \Gamma_1^{pt}(\Delta E) \right\} \;,
\label{eq:LAT-pt}
\end{eqnarray}
where $L$ is the linear extent of the lattice. As pointed out in~\cite{Carrasco:2015xwa}, the small momenta contributions to $\Gamma_0(L)$ and $\Gamma_0^{pt}(L)$ are the
same, hence the infrared divergences cancel in the difference. The same is true for the infinite volume combination $ \Gamma_0^{pt}+ \Gamma_1^{pt}(\Delta E)$,
therefore the two terms in the brackets on the r.h.s of eq.~\ref{eq:LAT-pt} are separately infrared finite (and, incidentally, gauge invariant) and have a well defined infinite volume limit. \\
The correlation functions needed in the  lattice computation of $\Gamma_0(L)$ and in particular
the three-point functions required for the non-factorizable terms
are explicitly constructed in~\cite{Carrasco:2015xwa}. The implementation of the method
is computationally demanding, but seems within reach of present resources, and first studies should soon be performed.
\section{Results and perspectives for  hadronic decays on the lattice}
Many phenomenologically interesting transitions involve hadronic two-body final states and the
lattice would be extremely useful in providing first-principle computations
which would serve in clarifying existing tensions (e.g., the one mentioned in the angular analysis of  $B^0 \to K^{*0}(K\pi) \mu^+ \mu^-$ by LHCb),
or give an ab-initio explanations of long-standing puzzles such as the $\Delta I=1/2$ rule and the value of $\varepsilon'/\varepsilon$ in $K \to \pi \pi$ decays. 
However, there is no simple relation among Euclidean correlators and the desired Minkowski-space transition matrix elements, a fact which is known as the 
``Maiani-Testa no-go theorem''~\cite{Maiani:1990ca}. A solution to this problem, for the case where
one two-particle state only (say, $\pi\pi$) is kinematically accessible or coupled to the initial state ($K$),  
was developed by L\"uscher and Lellouch in a series of papers~\cite{Luscher:1986pf,Luscher:1990ux,Luscher:1991cf,Lellouch:2000pv}.
In a first step a relation is established, in Minkowski-space, between the finite volume dependence of the 
energy levels of two-particle states ($\pi\pi$) and the infinite-volume S-matrix and phase shifts.
Since energy levels are directly computable in Euclidean-space, this allows to measure elements of the S-matrix 
on the lattice.
The kaon is introduced in a second step and it is coupled to the two-pion states 
through a small, perturbative, Weak-Hamiltonian term $H_W$. The lattice volume has to be tuned such that one of the
two-pion energy levels gets degenerate with the kaon, that is what is usually called ``matching the kinematics''.
At this point degenerate perturbation theory can be used, and as in the first step, a relation (in terms of
``Euclidean'' quantities) is established among
the perturbative corrections to the two-particle energy levels in finite volume
and the perturbative corrections to the infinite-volume S-matrix, which is to say a relation among the finite- and infinite-volume 
versions of the $\langle \pi\pi | H_W | K \rangle$ matrix element. The latter then gives the $K \to \pi\pi$ transition amplitude.

The approach has been generalized in~\cite{Hansen:2012tf,Briceno:2012yi,Briceno:2014uqa} to the case of multiple strongly-coupled decay channels
into two scalar particles and to the case of external currents injecting arbitrary four-momentum as well as angular momentum.
These are first steps towards lattice computations of amplitudes for processes such as $D \to \pi \pi$ and $D \to K\overline{K}$ and
towards study of meson decays as  $B^0 \to K^{*0}(K\pi) \mu^+ \mu^-$.

In the case of the $K \to \pi \pi$ transitions, numerical results became recently available with good control over all the systematics
including continuum limit extrapolations. These are outstanding results of many years of efforts and attempts.
In~\cite{Blum:2015ywa} the amplitude $A_2$ for a kaon to decay into two pions with isospin $I=2$ has been computed on two lattices  with resolutions 
$a=0.11$ fm and $a=0.084$ fm respectively. The calculations have been performed using 2+1 flavors of domain wall fermions with pions
at the physical mass and $L \approx 5$ fm. The matrix elements of three different operators have to be combined in this case and the final result, 
extrapolated to the continuum limit, reads
\begin{eqnarray}
{\rm Re} A_2 &=& 1.50(4)_{\rm stat}(14)_{\rm syst} \times 10^{-8} \; {\rm GeV}\;, \\
{\rm Im} A_2 &=&-6.99(20)_{\rm stat}(84)_{\rm syst} \times 10^{-13} \; {\rm GeV}\;,
\end{eqnarray}
which is well consistent with both the very accurate (but different) experimental numbers for ${\rm Re} A_2$ from charged and neutral kaon decays.
The error on the lattice value is dominated by systematics, in particular by the uncertainty in the perturbative
evaluation of the Wilson coefficients, currently known at NLO. 

The computation of the $A_0$ amplitude is much more demanding as 10 operators~\cite{Buchalla:1995vs}, 
including QCD penguins producing quark disconnected diagrams, need to be considered~\footnote{Actually 12 operators appear, but $Q_{11}$ and $Q_{12}$
are usually neglected at this level of accuracy since their contributions are suppressed by a factor $m_\pi^2/m_K^2$, as discussed in$\,^{25}$.}.
However, after the conference, two independent preliminary results~\cite{Bai:2015nea,Ishizuka:2015oja} appeared, both using a single lattice spacing and both reporting on
a computation of the $\Delta I=1/2$ amplitude $A_0$.
\section{Conclusions}
Flavor Physics is still playing a prominent role in the indirect search for New Physics. 
At the same time, and while finalizing the analysis of LHC run~I data, new signals from direct searches may emerge (as for example
in the search for resonances presented in~\cite{Aad:2015owa}, and interpreted within composite dynamics models in~\cite{Franzosi:2015zra}), 
which will hopefully be confirmed by run~II.

The picture provided here is obviously incomplete and the result of our taste and interests, but we hope to have given a 
flavor of the important role, the main challenges and the exciting future directions and perspectives for lattice gauge theories within Flavor Physics.
As the keywords seem to be precise and rare, the lattice community is tackling all subleading effects (e.g., isospin breaking) and theoretical
obstructions (e.g. multi-hadron decay channels) to give an indispensable contribution to the quest for New Physics.
\section*{Acknowledgments}
I thank the organizers of Moriond EW 2015 for the invitation to this very interesting and enjoyable conference. 
I am grateful to my colleagues in the FLAG collaboration for all their work in completing the corresponding review.
In particular I wish to thank the members of the Heavy Quark working groups, Aida X. El-Khadra, Yasumichi Aoki, Enrico Lunghi, Carlos Pena,
Junko Shigemitsu and Ruth Van de Water.
I thank Tadeusz Janowski, Amarjit Soni, Nazario Tantalo, Uli Haisch and Sebastian J\"ager for useful discussions and Francesco Sannino
for a critical reading of the manuscript.
This work was partially supported by 
the Spanish Minister of Education and Science, project RyC-2011-08557, and by 
the Danish National Research Foundation under the grant n. DNRF:90.
%
%
\vspace{-0.2cm}
\section*{References}
\end{document}